# AI-Supported Mini-Labs: Combining Smartphone-Based Experiments and Multimodal AI

Jochen Kuhn[1], David J. Rakestraw[2], Stefan Küchemann[1] & Patrik Vogt[1]
[1]LMU Munich, Faculty of Physics/Chair of Physics Education Research, 80539 Munich, Germany
[2]Lawrence Livermore National Laboratory, Livermore, CA 94550

Multimedia learning has always been a topic in physics education – whether through specialized digital data acquisition systems, simulations, or animations. Due to increasing digitalization, more and more media from our everyday lives have also found their way into education in recent decades.

**Smartphone-Based Mobile Mini-Labs**

Thus, for almost 15 years, the potential of smartphones and tablet PCs as powerful experimentation tools in physics education has been explored (Vogt, Müller & Kuhn, 2009; Kuhn & Vogt, 2012). These mobile devices have quietly evolved into highly sophisticated measurement platforms in science classrooms across the globe. With integrated sensors such as accelerometers, gyroscopes, LiDAR sensors, magnetometers, microphones, microwave antennas (WiFi, Bluetooth, and GPS), and multiple cameras, they are capable of recording a wide range of physical data. This turns them into portable "Mini-Labs," enabling students to conduct a wide range of highly sophisticated experiments using the technology they already carry with them. The use of these mobile mini-labs in physics education is closely linked to teaching-learning theories of situated and context-based learning, as well as the cognitive theory of multimedia learning (Kuhn & Vogt, 2022). Learners feel more engaged and autonomous when they use familiar everyday devices to explore physical phenomena – leading to higher motivation and an increased sense of responsibility for their own learning process. Furthermore, the devices have extensive computational power, high resolution touch screen displays, and audio/visual interfaces which support a growing number of mobile applications. These capabilities provide students with various pathways to engage with their experimental data, including diagrams, oscillograms, frequency spectra, etc., in addition to text and tables (for various examples, see Kuhn & Vogt 2022). However, the opportunities to record and analyze everyday phenomena with these powerful mini-labs are sometimes limited in regular classes, because everyday phenomena can be quite complex, and the necessary analytical methods are often not available in school curricula.

**Multimodal Large Language Models**

This limitation can be increasingly overcome thanks to rapid advances in the field of large language models which can serve as co-investigators and greatly expand opportunities for student exploration. Since the release of GPT-4o in mid-2024, many of these foundation models are capable of processing and generating multimodal content (Multimodal Large Language Models, MLLM) significantly expanding their use cases. Thus, such generative AI tools can interpret and generate images, audio, and video in addition to text. They thereby offer new possibilities to support learners in processing complex data. These models can function as intelligent tutors, helping learners optimize experiment designs, analyze and visualize experimental results, interpret charts and tables, explain underlying concepts, and even guide the construction of scientific arguments (Bewersdorff et al., 2025). What distinguishes MLLMs from conventional AI systems is their ability to work across modalities. They enable learners to switch fluently between text-based, visual, and symbolic representations of knowledge. This capability aligns with the cognitive theory of multimedia learning, which emphasizes that integrating multiple forms of representation can promote understanding and reduce cognitive load.

**MLLM-Supported Analysis of Mobile Mini-Labs**

The AI-based enhancement of experimenting with smartphones thus also enables the analysis of complex everyday phenomena in regular classes at secondary levels 1 and 2 (first examples include Klay et al., 2025; Vogt et al, 2025a; Vogt et al., 2025b). With such AI-Augmented Pocket Labs, learners can individually explain phenomena scientifically, create and evaluate designs for scientific investigations, and critically interpret scientific data and evidence. These competencies correspond to the PISA 2025 Science Framework.

This article illustrates this concept idea using three examples using different sensors and demonstrating analysis of different data modalities.

**Example 1: Determination of the drag coefficient**

The determination of a vehicle's drag coefficient using the accelerometer integrated into a smartphone was first published in 2015 (Fahsl et al., 2015; for an English-language contribution, see Fahsl & Vogt, 2018). The educational potential of this experiment unfolds on multiple levels: students collect the measurement data independently, using a familiar and personal device in a real-world context. This approach strongly aligns with the widely advocated call for greater contextualization of physics content in the spirit of situated learning.

Furthermore, students physically experience the different phases of driving—acceleration with gear shifts and coasting—which makes the activity tangible and personal. Their subjective experience of motion can then be related to the objective analysis of measured data, thus fostering and incorporating elements of embodied cognition (Vogt et al., 2016). The connection between acceleration and velocity through a time-based comparison offers numerous opportunities for discussion and supports the development of a solid conceptual understanding of kinematics.

Not least, the data allows for an estimation of the vehicle's drag coefficient and rolling resistance coefficient—quantities that are otherwise difficult to access in a school setting. However, determining the velocity from acceleration data requires numerical integration. Since this computational method is typically not covered in basic calculus classes, it would need to be taught by the physics instructor. This can be done using a spreadsheet or a variety of other computational tools but takes significant time and adds additional cognitive load to the investigation.

With the advent of multimodal large language models (MLLMs), students can now independently generate and evaluate diagrams without requiring advanced mathematical knowledge. For initial AI-assisted data analyses, it is advisable to proceed step by step in dialogue with the model. This allows students to understand the analysis process and simultaneously gain insight into how the AI works. Building on these experiences, more structured prompts—easily adaptable to different contexts—can be used for further analyses. We have opted for a middle ground: starting with a more comprehensive prompt (see Fig. 1; followed by the system answer of ChatGPT in Fig. 2), then refining the evaluation of the experiment through a more focused and interactive dialogue (see Fig. 3-Fig. 10).

*Input:* I conducted a physics experiment and need your help analyzing the data. I've uploaded the data as an Excel file.

**Brief Description of the Experiment**

The uploaded dataset contains the acceleration profile of a vehicle that accelerated from a standstill to its maximum speed. The acceleration in the direction of motion was measured. Once the maximum speed was reached, the vehicle was put into neutral and allowed to coast on a level road.

**Graphical Analysis**
– Determine the velocity as a function of time based on the acceleration data.
– Create a plot showing acceleration as a function of time *t*, covering the entire measurement period.
– Do not include decimal places in the axis tick labels.
– Do not plot the raw acceleration data, but instead a heavily smoothed version in red.
– Plot the velocity in the same diagram and add an additional vertical axis on the right for it. Use "Velocity" (not "Speed") for the label.
– Plot the velocity curve in blue.
– Always use SI units.

**General Formatting Guidelines for the Diagram**
– Do not include a title.
– Label the axes with the physical quantity in words (not the symbol) followed by the unit symbol in parentheses. For example, for the *x*-axis: Time (s).
– Use a step size of 1 m/s² on the acceleration axis, 5 m/s on the velocity axis, and 10 s on the time axis.
– The first data point should be adjusted to the time *t* = 0 s.

*Fig. 1: Comprehensive prompt to analyze the data of an acceleration process of a vehicle*

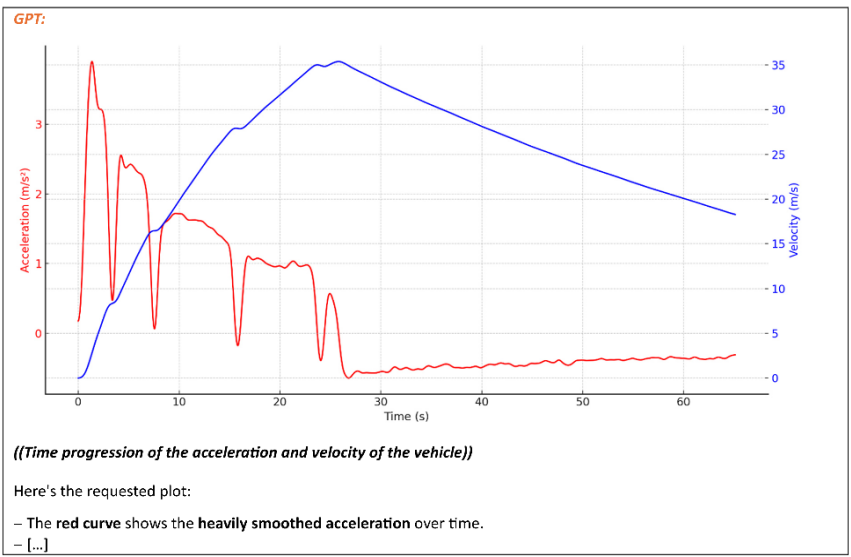

*GPT:*

*((Time progression of the acceleration and velocity of the vehicle))*

Here's the requested plot:

– The **red curve** shows the **heavily smoothed acceleration** over time.
– [...]

*Fig. 2: System answer of ChatGPT to prompt in Fig. 1: Time progression of the acceleration and velocity of the vehicle*

*Input:* As soon as the vehicle reaches its maximum velocity, it begins to coast. For the coasting phase only (starting from 30 s), create a plot of the heavily smoothed acceleration values as a function of the square of the velocity, using red dots. Do not connect the points with lines. Use the magnitude of the acceleration values (ignore the sign). Additionally, add a blue regression line and extend it all the way to the y-axis so the y-intercept is visible.

What is the coefficient of determination of the regression? What are the values of the y-intercept and the slope of the regression line? Provide the correct physical units for all these parameters. Do not display the regression equation in the plot.

*Fig. 3: Stepwise refining prompt 1 for a plot with smoothed data and regression equation*

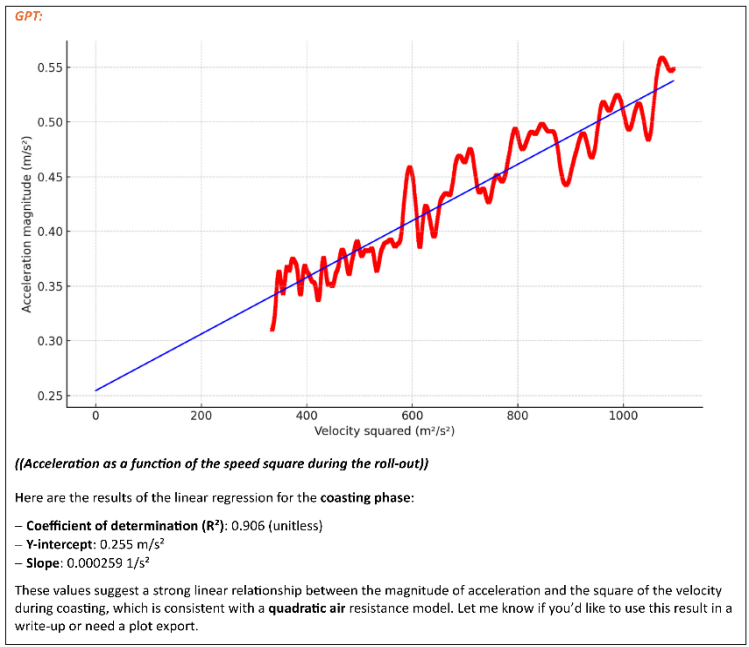

*Fig. 4: System answer of ChatGPT to prompt in Fig. 3*

**Input:** Use the regression parameters to calculate the rolling resistance coefficient and the drag coefficient. Take the following values into account: frontal area of the vehicle A = 2.22 m², mass m = 1430 kg, air density ρ = 1.2 kg/m³. Show all steps in the calculations clearly. Always include physical units, not just in the final results.

*Fig. 5: Stepwise refining prompt 2 to calculate the rolling resistance coefficient and the drag coefficient*

**GPT:** To compute the rolling resistance coefficient $c_r$ and the drag coefficient $c_d$, we start by analyzing the total acceleration during the coasting phase. The total deceleration $a$ (magnitude of acceleration) arises from:

$$a = a_{\text{roll}} + a_{\text{drag}} = \frac{F_{\text{roll}}}{m} + \frac{F_{\text{drag}}}{m}$$

Where:
- $F_{\text{roll}} = c_r \cdot m \cdot g$
- $F_{\text{drag}} = \frac{1}{2} \cdot \rho \cdot A \cdot c_d \cdot v^2$

So:

$$a = \frac{c_r \cdot m \cdot g}{m} + \frac{\frac{1}{2} \cdot \rho \cdot A \cdot c_d \cdot v^2}{m}$$

This simplifies to:

$$a = c_r \cdot g + \left(\frac{1}{2} \cdot \frac{\rho \cdot A}{m} \cdot c_d\right) \cdot v^2$$

Comparing with the linear regression equation:

$$a = \text{intercept} + \text{slope} \cdot v^2$$

we identify:
- $c_r \cdot g = \text{intersept} \Rightarrow c_r = \frac{\text{intercept}}{g}$
- $\frac{1}{2} \cdot \frac{\rho \cdot A}{m} \cdot c_d = \text{slope} \Rightarrow c_d = \frac{2 \cdot m \cdot \text{slope}}{\rho \cdot A}$ [...]

**1. Rolling Resistance Coefficient $c_r$:**

$$c_r = \frac{0.255 \text{ m/s}^2}{9.81 \text{ m/s}^2} = 0.026$$

Unit: dimensionless

**2. Drag Coefficient $c_d$:**

$$c_d = \frac{2 \cdot 1430 \text{ kg} \cdot 0.0002591 \text{ 1/s}^2}{1.2 \frac{\text{kg}}{\text{m}^3} \cdot 2.22 \, m^2} \approx 0.278$$

Unit: dimensionless

**Final Results**
- **Rolling resistance coefficient** $c_r \approx 0.026$
- **Drag coefficient** $c_d \approx 0.278$

Let me know if you'd like a formatted version for a report or further interpretation of the results.

*Fig. 6: System answer of ChatGPT to prompt in Fig. 5*

> *Input:* Compare the results with values from the literature and evaluate the experimental outcome.

Fig. 7: Stepwise refining prompt 3 to compare the results with literature values

> **GPT:** To evaluate the experimental outcome, let's first compare your calculated values to those typically found in the literature:
>
> **1. Rolling Resistance Coefficient $c_r$**
>
> Your result: $c_r = 0.026$
>
> **Typical values from literature:**
> - Passenger cars: $c_r = 0.01$ to $0.015$ on asphalt for properly inflated tires.
> - $c_r = 0.02$ or higher is possible with underinflated tires, rough surfaces, or aging components.
>
> **Evaluation:**
> Your measured value of $c_r = 0.026$ is **somewhat higher than expected**, which could be due to:
> - Increased mechanical resistance in the drivetrain.
> - Tire pressure not being optimal.
> - A rougher road surface.
> - Simplifying assumptions (e.g., neglecting wind or slope effects).
>
> This suggests **either a real-world deviation** from ideal conditions or an accumulation of small experimental uncertainties.
>
> **2. Drag Coefficient $c_d$**
>
> Your result: $c_d \approx 0.278$
>
> **Typical values from literature:**
> - Modern sedans: $c_d = 0.25$ to $0.30$
> - Sports cars: $c_d < 0.30$
> - SUVs or vans: $c_d = 0.35$ to $0.45$
>
> **Evaluation:**
> Your measured $c_d$ falls well within the expected range for a **standard passenger car**.
> The result indicates that **the aerodynamic estimation is quite reasonable**, suggesting good data quality and a successful application of the coasting method.
>
> **Overall Assessment**
> - The **drag coefficient** measurement is **highly plausible** and supports the quality of your smoothed data and regression approach.
> - The **rolling resistance coefficient** is **somewhat overestimated**, but not unreasonably so—likely reflecting realistic road and vehicle conditions rather than idealized ones.
> - The experiment appears to have been **successful**, especially in identifying and separating quadratic air resistance from constant rolling resistance effects.
>
> Would you like a written summary of this evaluation in a format suitable for a lab report or presentation slide?

Fig. 8: System answer of ChatGPT to prompt in Fig. 7

> *Input:* I have an idea to explain the high rolling resistance coefficient. The model assumes that only the rolling resistance force is independent of velocity. But doesn't this also apply to the friction in the wheel hub? This friction component is also included in the rolling resistance force in our approach. What do you think?

Fig. 9: Stepwise refining prompt 4 to explain the high rolling resistance coefficient

> **GPT:** Yes, you're absolutely right—and that's a very insightful observation!
>
> In your model, the **rolling resistance force** is lumped into a single term that's **velocity-independent**, and it's described using the simplified formula:
>
> $$F_{\text{roll}} = c_r \cdot m \cdot g$$
>
> This term is meant to include **all velocity-independent resistive forces**, and not just the classical **tire deformation** losses (which are typically meant by "rolling resistance" in isolation).
>
> But in practice, especially in a real vehicle, this also includes:
>
> - **Wheel hub bearing friction**
> - **Brake drag** (slight friction even when not applied)
> - **Drivetrain friction** (if any residual coupling remains)
> - Possibly **internal friction** in the tires or other components that don't scale with velocity
>
> These components contribute **additively** to the deceleration that is independent of velocity. So yes:
>
> **The wheel hub friction is conceptually included in the rolling resistance force in your model.**
>
> That would explain a higher apparent $c_r$, since your fitted intercept includes all of those effects.
>
> **Bonus Thought**
>
> This highlights how important it is to distinguish between:
>
> - **Rolling resistance coefficient** $c_r$ in a strict tire-physics sense vs.
> - **Effective or lumped** $c_r$ used in simplified modeling like your experiment.
>
> The latter is more practical in experimental physics teaching, but it's good to be aware of what hides inside that lumped term. [...]

Fig. 10: System answer of ChatGPT to prompt in Fig. 9

The conventional data analysis of the same dataset using spreadsheet software aligns well with the results obtained from the MLLM-based analyses.

**Example 2: Measurement of the ionospheric height**

In addition to oscillograms and frequency spectra (Vogt et al., 2025a; Vogt et al., 2025b), spectrograms also enable the graphical representation and quantitative analysis of acoustic signals. A less obvious application example involves lightning discharges, which emit electromagnetic signals in the Very Low Frequency (VLF) range between 3 and 30 kHz. To analyze these signals using ChatGPT-4o, they must first be recorded using a simple receiving device functioning as an acoustic transducer (Vogt & Müller, 2014) and saved as a WAV file on a computer. An introduction to lightning-generated atmospherics—or "sferics" for short—is either embedded in the prompt or can be found in detail in the referenced publication.

> *Input:* I am a physicist and would like to investigate a phenomenon of atmospheric electrodynamics. You are to assist me in the analysis of an audio file, which I have uploaded as a WAV file.
>
> **Brief description of the experiment**
> The discharge channel of a lightning strike represents a huge, branched antenna structure extending over hundreds to thousands of meters, in which an impulsive electric current flows. It consists of preliminary, main, and intermediate discharges. The signals emitted during this process are referred to as atmospherics, or simply "sferics." They have a very broadband spectrum, with detectable frequencies ranging from 1 to $10^9$ Hz. However, the majority of the energy is concentrated in the VLF range.
>
> If atmospheric disturbances are recorded with a longwave receiver, and the signal propagates via the ground wave, a short crackling sound is heard during acoustic conversion; all frequencies arrive at the receiver at the same time. In the spectrogram—which plots frequency over time—this corresponds to a straight vertical line.
>
> However, similar to the electromagnetic waves of traditional medium-wave radio, sferics can also propagate via the sky wave, i.e., through multiple reflections in the natural waveguide formed by the Earth and ionosphere. To describe this waveguide quantitatively, we can approximate it using the model of a planar waveguide. This consists of two ideally conducting, parallel plates with a vacuum in between. In such a waveguide, only electromagnetic waves with wavelengths up to twice the plate separation $h$ can propagate. Below the resulting cut-off frequency $f_c$, exponential attenuation occurs; above it, free propagation is possible.
>
> In this model, the propagation speed of an electromagnetic signal depends on its frequency. Near the cut-off frequency of the waveguide, the signal speed decreases significantly; at higher frequencies, it approaches the speed of light in a vacuum. As a result, the broadband signal of a sferic undergoes dispersion, which acoustically manifests as a chirping sound. In the spectrogram, the vertical line of the sferic becomes a characteristic "hook" between 1.5 and 2 kHz, since lower-frequency components now arrive at later times. Such a signal is called a **tweek**.
>
> **Representation in the spectrogram**
> The uploaded WAV file contains a sequence of tweeks. Display them in the form of a spectrogram. Choose the color scale so that the signals—particularly their characteristic hook near the cut-off frequency—have high contrast against the noise. You can refer to the uploaded image to see the typical appearance of a tweek in a spectrogram. Do not include a title in the diagram.
>
> **Estimation of ionospheric height**
> From the spectrogram, determine the cut-off frequency of the tweeks and use this measurement to calculate the corresponding reflection height (ionospheric height). Provide a detailed derivation of your calculation, including physical units throughout. Present the final result in kilometers. Compare the calculated height to values from the literature (ionospheric height at night after the disappearance of the D-layer) and assess the outcome of the measurement.

*Fig. 11: Prompt to analyze the data of an audio file to investigate phenomena of atmospheric electrodynamics*

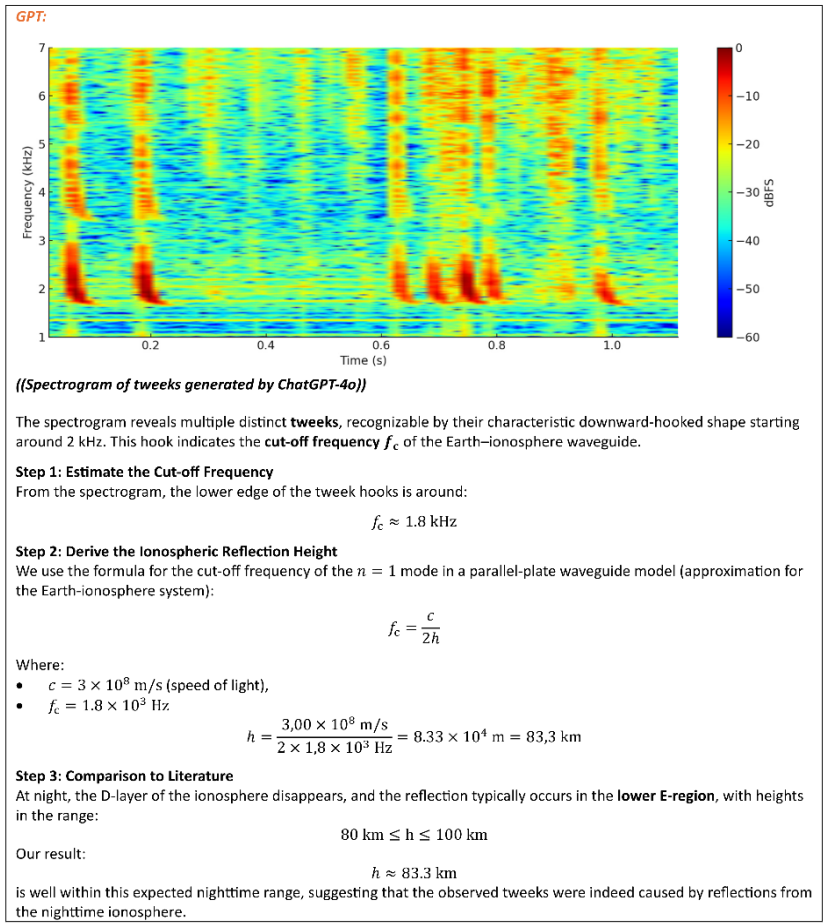

*Fig. 12: System answer of ChatGPT to prompt in Fig. 11: Spectrogram of tweeks*

This example illustrates that an MLLM is also capable of handling alternative data formats—in this case, an audio file in WAV format. Moreover, it demonstrates the remarkable precision with which acoustic analyses can be conducted (see also Vogt et al., 2025a; Vogt et al., 2025b).

**Example 3: Real-time spectroscopy of blood volume dynamics**

In this example we analyze a video that has recorded the transmission of light through a student's finger using the smartphones CMOS image sensor. The data was collected by placing their finger over the front camera and measuring the transmission. While each CMOS pixel only has a dynamic range of 256, the integration of more than a million pixels can provide extraordinary precision in measuring small changes in transmission. Measuring the transmitted intensity frame-by-frame allows for time resolved spectroscopy of blood volume in a finger during the cardiac cycle. The phototherapeutic window between 650-850 nm makes the measurement of the red pixels ideal (Rakestraw, 2025).

> *Input:* The uploaded dataset contains a video that captures the light being transmitted through a finger placed over the front camera of a smartphone. This is a simple spectroscopy experiment that uses the ambient light as the light source and the CMOS image sensor as the detector. In this experiment, the entire array of individual diodes in the image sensor can serve one integrated detector with 3 different colored filter options. The amount of light transmitted should change with the variation in blood volume in the light path through the finger, reflecting properties of an individual's cardiac cycle and circulatory system.
>
> **Analysis**
> - Integrate all of the red pixels in the image and calculate the average intensity of the light for each frame of the video.
> - The change of the intensity over each cardiac cycle will be small compared to the overall light intensity reaching the image sensor. Apply a high pass filter to remove the DC component of the averaged intensity.
> - Invert the data and then plot the intensity vs time. This will result in a positive change associated with an increase in blood volume.
>
> **Graphing**
> Plot the data using a scatter plot

*Fig. 13: Prompt to analyze the data of a video file to investigate transmission of light through a student's finger*

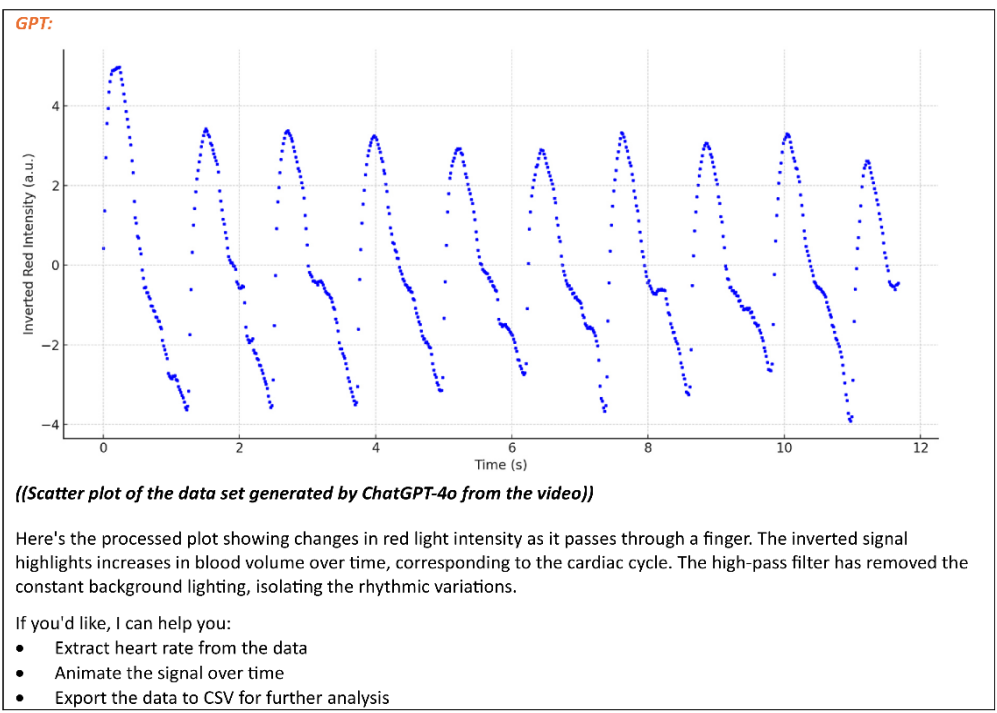

*((Scatter plot of the data set generated by ChatGPT-4o from the video))*

Here's the processed plot showing changes in red light intensity as it passes through a finger. The inverted signal highlights increases in blood volume over time, corresponding to the cardiac cycle. The high-pass filter has removed the constant background lighting, isolating the rhythmic variations.

If you'd like, I can help you:
- Extract heart rate from the data
- Animate the signal over time
- Export the data to CSV for further analysis

*Fig. 14: System answer of ChatGPT to prompt in Fig. 13: Scatter plot of the data set*

The analysis by ChatGPT-4o was very fast and dramatically enhances the opportunity for doing this type of analysis in the classroom. The rise and fall of the absorption peaks for each cardiac cycle are rich with physiological information. The quick analysis process with AI allows students to design interesting physiological experiments and study the response. Even a single experiment is difficult to perform without the assistance of AI as students need to gain access and learn how to use video analysis software distracting their focus from the interesting physics and biology.

**Advantages Over Conventional Analysis Methods**

The tasks performed by the MLLM in the examples presented, along with the resulting advantages over conventional data analysis methods, are summarized in the following overview.

| Example | Task Performed by the MLLM | Advantages Compared to Conventional Methods |
|---|---|---|
| Determination of the drag coefficient | Graphical representation | Time savings; high-quality visualizations; exportable as pixel or vector graphics; we have given the relationship $a \sim v^2$, this dependency could also have been found very quickly by the students with the help of ChatGPT |
| | Smoothing of data series | Generally not part of school-level mathematics; time savings |
| | Numerical integration | Generally not part of school-level mathematics; time savings; the independent creation of diagrams is only made possible by the use of the MLLM; stepwise explanation of analysis |
| | Calculation and graphical representation of best-fit lines | Time savings, no additional software required |
| | Step-by-step derivation of calculation formulas | ChatGPT-4o can take on the role of a tutor |
| | Calculation of the desired quantities | Time savings |
| | Comparison with literature values and evaluation | Time savings and discussion partner |
| Measurement of the ionospheric height | Analysis of a WAV file and generation of a spectrogram | High-quality visualizations; scaling, axis labeling, and color design can be individually customized; most apps only allow for screenshots, whereas MLLM-generated images can be exported as high-resolution PNG files or even as vector graphics (e.g., in PDF format). |
| | Determination of the cutoff frequency | High accuracy |
| | Calculation of the ionosphere height and comparison with literature values | Time-savings; if necessary, explanation of the calculation steps; execution of the calculation; comparison with literature values |
| | Other | Further analyses are also possible, such as the representation of an oscillogram, a frequency spectrum, and much more. |
| Real-time spectroscopy | Data analysis | Completely automated; time-saving; no additional software or training required |
| | Graphical representation | Time savings; high-quality visualizations |

*Tab. 1: Advantages of MLLM-supported data analysis in the examples presented*

One major advantage of the tasks performed by the MLLM is that students only need basic knowledge of prompt engineering—without having to learn additional software tools or complex computational methods. This leaves more room in the classroom for engaging with the underlying physical concepts.

**Discussion**

Not least, the three examples show that the combination of smartphone experiments and MLLMs offers a beneficial approach to capturing and analyzing scientific data, as well as critically questioning evidence. A key advantage of this approach lies in fostering representational competence – the ability to navigate and translate between different forms of scientific information representation, such as diagrams, graphs, formulas, and written explanations. Studies show that this competence is crucial for learning success in the sciences (e.g., Treagust, Duit & Fischer, 2017), yet many students struggle with it (Scheid et al., 2017). MLLMs can function here as dynamic learning aids by breaking down complex visual representations or generating meaningful representations from raw data, thereby promoting representational competence (Klein, Müller & Kuhn, 2017) – doing so in an individualized and differentiated manner. For example, they can convert 3 dimensional datainto a visually informative diagram (e.g., spectrogram), highlight relevant features, and explain their meaning – depending on the learners' prior knowledge – in simple language or at a higher technical level. In this context, the ability of MLLMs to analyze sensor data (whether from interfaces or smartphones), audio files, images and videos should be emphasized once again.

The integration of smartphones and MLLMs aligns well with fostering 21st-century skills such as problem-solving, creativity, collaboration, and digital literacy. Learners are encouraged not only to engage with physical phenomena and processes but also to interact meaningfully with data, models, and representations. Through the critical interpretation of evidence – e.g., by analyzing data trends, questioning irregularities, or considering experimental errors – learners develop an inquiring and investigative attitude, which is a key feature of science education.

Despite all these advantages, it is crucial to view the integration of AI in education from a critical perspective. Concerns regarding data privacy, ethical use, and so-called AI hallucinations (i.e., generating plausible but false information) must be addressed through clear guidelines and measures to promote digital literacy. This also applies to the AI-Augmented Pocket Labs approach described here. Educators should find a careful balance – MLLMs should support learning, not replace it. Students should be encouraged to use AI tools as thinking partners, not as answer machines. Furthermore, findings on the learning impact of generative AI systems are currently still very contradictory. For instance, some review studies report positive learning effects when using such systems (e.g., Deng et al., 2025). However, the results of these studies are quite controversial due to methodological weaknesses. Furthermore, other contributions point to negative effects when using such tools for learning (see, e.g., Stadler, Bannert & Sailer, 2024; Krupp et al., 2024a, b; Fan et al., 2024), whereas studies on learning with smartphone experiments alone are coherently positive (Hochberg, Kuhn & Müller, 2018; Hochberg et al., 2020; Becker et al., 2020a, b; Laumann et al., 2024).

In summary, the combination of the smartphone as an experimental tool and generative AI as an intelligent, multimodal assistance system opens up a promising approach for individualized learning in physics education. Besides the conceptually obvious opportunities, however, it is still necessary to provide scientific evidence for the positive learning effects of this approach in physics education. However, in an educational landscape increasingly shaped by AI, such hybrid models can play a key role in preparing students not just for learning, but for practical application – to be curious, critical, and creative.